\documentclass[aps,prl,groupedaddress,showpacs,letterpaper,preprintnumbers,amsmath,amssymb,twocolumn]{revtex4}
\usepackage{graphicx,amsmath}

\begin{document}
\title{Geometric mesoscopic correlations in quasi-one dimension}
\author{Alexey Yamilov}
\affiliation{Department of Physics, University of Missouri-Rolla, Rolla, MO 65409}
\email{yamilov@umr.edu}
\date{\today}

\begin{abstract}
We study analytically and numerically field/intensity correlations in wave transport through volume-disordered waveguide. The obtained channel and spacial correlations deviate from those found in framework of Dorokhov-Mello-Pereyra-Kumar (DMPK) formalism, that we relate to inapplicability of equivalent channel approximation in DMPK. We show that this can be remedied by introducing boundary correction -- an escape function which depends on the waveguide geometry -- that describes wave transport near a boundary between random medium and free space. We obtain the expressions for field/intensity channel and spacial correlation functions which agree with the numerics and are consistent with the perturbative expressions in slab geometry as well as experiments conducted in Q1D.
\end{abstract}
\maketitle

Mesoscopic fluctuations, such as universal conductance fluctuations \cite{altshuler}, are rooted in nonlocal correlations \cite{nonlocal,shapiro_nonlocal} that appear due to the interference effects when a wave undergoes multiple scattering events in random medium. Although common to both the electron and the electromagnetic waves, the experiments with the light give more detailed information about transport \cite{feng_phys_rep,rossum_rmp}, for example mesoscopic correlation of optical fields leads to readily observable speckles \cite{sebbah_book}.

The universality of the statistics of wave transport and ability to study localization-delocalization transition as a function of only system length, made Q1D geometry ({\it i.e.} a wire) a fruitful testbed for studying mesoscopic phenomena \cite{mirlin_g,mello_book,beenakker_rmp}. Due to quantization of the transverse momentum, the wave transport can be described in terms of $T_{ab}=t_{ab}t_{ab}^{*}$ -- the channel transmission coefficient -- that measures the flux transmitted in the channel $b$ when a unit flux enters via channel $a$.  Summation over the channel indexes gives the transmission $T_a=\Sigma_b T_{ab}$ and the dimensionless conductance \cite{landauer_formula}:
\begin{equation}
T=\displaystyle\sum_{ab}T_{ab}=\displaystyle\sum_{ab} k_a k_b \mathcal{T}_{ab}; \ \ \ \ \ g\equiv\langle T\rangle.
\label{g}
\end{equation}
where $k_{a,b}$ are the longitudinal component of the momentum, necessary for the proper \cite{landauer_formula_stone} treatment of incoming/outgoing fluxes; $\mathcal{T}_{ab}$ is the squared amplitude of the mode $b$. For the light, $T_{ab},\ T_a$ and $T$ are directly measurable\cite{marin_total_transm,azi_nature}.

Diagrammatic\cite{altshuler} and Dorokhov-Mello-Pereyra-Kumar (DMPK) \cite{dorokhov,mpk} techniques were highly successful in studying mesoscopic phenomena and were applied to investigate the correlations in the slab \cite{feng_kane_lee_stone} and Q1D \cite{mello_correlations} geometries respectively. DMPK is a self-embedding approach that assumes the incremental transfer matrices to be random with only time reversal and energy conservation being the constraints. Such random matrix theory (RMT) relies on isotropy assumption, which leads to a mathematically convenient anzatz \cite{mello_book} but lacks a microscopical foundation \cite{tomsovic,tartakovski}. This approximation yields results which depend on number of channels $N$, irrespective of geometry of the waveguide. It can be recast into the equivalent channel approximation (ECA), validity of which was questioned before \cite{tomsovic,tartakovski,garcia1996} and recently received a renewed focus\cite{mello_2007}. ECA assumes that after scattering in a macroscopically thin slice of the disordered medium all channels become completely mixed and, thus, are statistically equivalent. This approximation is linked to DMPK's inability  \cite{tomsovic,mello_2007} to adequately describe diffusion in the direction transverse to the axis of the wire, as well as a discrepancy between mean free path (MFP) $\ell$ values in RMT and the transport theory \cite{beenakker_rmp,tartakovski}. Modified DMPK formalism, free of ECA, yields \cite{mello_2007} complex equations which resist an analytical treatment.

The purpose of this work is the detailed test of RMT$+$ECA predictions in Q1D \cite{mello_correlations,beenakker_rmp} with the emphasis on mesoscopic correlations. Using the direct numerical simulations, we obtain the transmission matrix without making ECA or any other approximations. We show that in the $N\rightarrow\infty$ limit, it is the quantity $t_{ab}\rho_a^{-1/2}\rho_b^{-1/2}$ that exhibits the properties of RMT transfer matrix: (i) its elements are identically distributed for different channel indexes $a,b$; (ii) field and intensity correlation functions match those derived in RMT. For $L\gg\ell$, the escape function $\rho_a$ depends only on the {\it geometry} of the waveguide, independent of its length $L$. This suggests that the limitations of ECA can be remedied by the multiplicative correction $\rho_a^{1/2}\rho_b^{1/2}$ to $t_{ab}^{(RMT)}$. We show that the resultant spacial correlations adequately describe the diffusion in the crossection of the waveguide and are consistent with diffusion theory.

{\it Numerical method:} 
In our numerical simulation, we consider 2D waveguide filled with random medium\cite{pofg,corr_passive}. We use finite difference time domain method to calculate the response of our system to pulsed excitation, followed by Fourier transformation which gives us the desired continuous-wave response\cite{corr_passive} so that $t_{ab}$ can be computed \cite{pofg}. Because the studied system is Q1D, the samples with values of $g$ in the range $\{0.4-4\}$ were obtained by varying the lengths $L$ of the random medium. To ensure that the obtained results do not depend on the microscopic structure of disorder, we operate in the regime of locally weak disorder, $k\ell\gg 1$. To analyze statistics of mesoscopic transport, ensembles of $10^4-10^5$ random realizations were obtained. 

\begin{figure}
\vskip -0.0cm
\centerline{\rotatebox{-90}{\scalebox{0.3}{\includegraphics{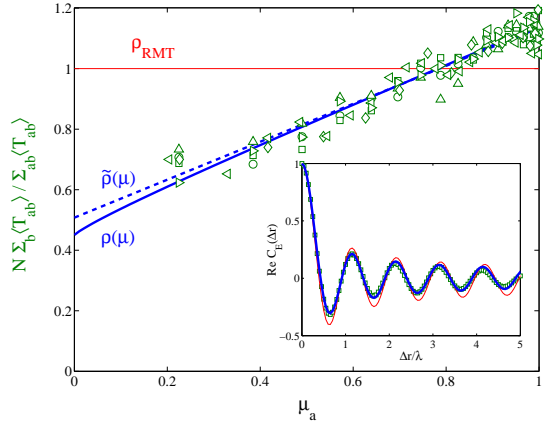}}}}
\vskip -0.5cm
\caption{\label{fig_rho}(a) Mean value of the transmission normalized by conductance $N\sum_b\langle T_{ab}\rangle/\sum_{ab}\langle T_{ab}\rangle$ obtained from numerical simulations in samples with different $N$, $g$, and $\ell$: 
(i)   ${\rm O}$\   -- $N=19,\ g=3.5$; 
(ii)  $\square$\   -- $N=19,\ g=1.5$; 
(iii) $\triangle$\ -- $N=19,\ g=0.5$; 
(iv)  $\Diamond$\  -- $N=15,\ g=0.8$; 
(v)   $\lhd$\      -- $N=28,\ g=3.0$; 
(vi)  $\rhd$\      -- $N=19,\ g=4.5$,  $\ell$ which is twice as long as in the previous  samples. 
The data agrees with  $\rho(\mu)$ and ${\tilde\rho}(\mu,\Delta=0.818)$ (thick solid and dashed lines) without any adjustable parameters. Thin solid line (a constant) depicts the Equivalent Channel Approximation made in DMPK approach. The inset shows spacial field correlation function (see Eq. \ref{Ce}) obtained in numerical simulation (squares), Eq. (\ref{Ce}) (thick line) and an expectation based on ECA (thin line). }
\end{figure}

{\it Validity of ECA:} 
We directly verify the earlier reports \cite{tomsovic,tartakovski,garcia1996,mello_2007} that $\langle T_{ab}\rangle$ has strong channel dependence. Fig. \ref{fig_rho} shows that it is described by the Chandrasekhar function as found for a slab\cite{rossum_rmp,tartakovski}
\begin{eqnarray}
\langle T_{ab}\rangle &=&\rho(k_a/k)\times\rho(k_b/k)\times\langle T\rangle/N^2\label{rho}\\
\rho\left(\mu\right)&=&C\ \exp\left[
-\frac{\mu}{\pi}\int_{0}^{\pi/2}
\frac{\ln\left( 1-\cos\beta (\beta/\sin\beta)^{D-2} \right)}{\cos^2\beta+\mu^2\sin^2\beta}     
\right], \nonumber
\end{eqnarray}
where $D=2,3$ is the dimensionality of the waveguide, and $C$ is chosen so that $\rho_a$ is normalized $\sum_a \rho_a=N$. Throughout this work $\langle ...\rangle$ imply averaging only over disorder configurations for given channel indexes $\{a,b\}$. The agreement between Eq. (\ref{rho}) and the numerical data in Fig. \ref{fig_rho} is achieved {\it with no adjustable parameters} for samples with different $N$, $g$, and $\ell$ with the only condition $L\gg\ell$ to avoid the ballistic regime \cite{tartakovski}. The deviations from $\rho$ in Fig. \ref{fig_rho} are attributed to limited statistics and finite number of channels $N$.

Such significant channel dispersion should not come as a surprise as it reflects fundamental wave coherence properties \cite{wolf_coherence}, and has a long history in radiative transfer theory (RTT) \cite{chandrasekhar}. For slab geometry ${\tilde \rho}(\mu)\propto \mu+\Delta$ is known to be a good approximation with $\Delta=z_b/\ell$ is the so-called extrapolation length. $\Delta$ equals to $0.818$ and $0.710$ (in 2D and 3D respectively) arise in RTT, whereas $\Delta_d=\pi/4\approx 0.785$ and $2/3\approx 0.667$ are obtained in the diffusion approximation \cite{rossum_rmp}. Also, the persistent channel dependence of $\langle T_a\rangle$ was seen in the simulations of Ref. \cite{markos_idist}, but was interpreted as a finite $N$ effect.

\begin{figure}
\vskip -0.2cm
\centerline{\rotatebox{-90}{\scalebox{0.25}{\includegraphics{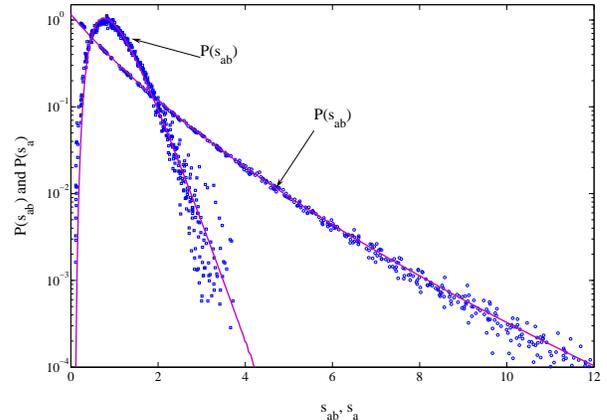}}}}
\vskip -0.3cm
\caption{\label{pofi} Statistical equivalence of the {\it normalized} transmission coefficients. Depicted are two groups of $a=[1..N]$ histograms $P(s_{ab}\equiv T_{ab}/\langle T_{ab}\rangle )$ ($b$ is set to $1$ for example) and $P(s_{a}\equiv T_{a}/\langle T_{a}\rangle )$ for the sample with $g=3.5,\  N=19$. Both groups are successfully fitted by the RMT's expression from Ref. \cite{kogan_formula}.}
\end{figure}

{\it Statistics of transmission coefficients:} 
RMT's relation $\langle T_{ab}\rangle=\langle T\rangle/N^2$ stems from the assumption of the statistical equivalence of all channels -- ECA. In view of Eq. (\ref{rho}), this connection is no longer a trivial one. {\it E.g.} in the surface disordered waveguides, strong channel dependence of MFP can even lead to coexistence of diffusive and localized regimes \cite{freilikher_coexistence} in different channels of the same sample. In contrast, in case of volume-disorder, considered here, the disorder-induced channel mixing should lead to equilibration for $L\gg\ell$. Independence of $\rho_a$ from $g$ and $L$ \cite{tartakovski} may be considered a manifestation of this effect.

Our simulations, Fig. \ref{pofi}, demonstrate that all channels indeed become statistically equivalent when the transmission coefficients are normalized by their averages -- $P(s_{ab}\equiv T_{ab}/\langle T_{ab}\rangle )$ and $P(s_{a}\equiv T_{a}/\langle T_{a}\rangle )$ are independent of channel indexes. It suggests that removing $\rho_{a,b}$ factor is sufficient to enforce ECA. It is, thus, tempting to associate $t_{ab}/\rho_a^{1/2}/\rho_b^{1/2}$ or $s_{ab}$ with $t_{ab}^{(RMT)}$ from RMT. However, $s_{ab}$ are no longer required to obey the composition rules
\begin{equation}
s_{a}\neq \displaystyle\frac{1}{N}\sum_b s_{ab};\ \ \  s\neq \displaystyle\frac{1}{N}\sum_a s_{a}.
\label{sa}
\end{equation}
The discrepancy appears due to correlations in different channels, to be discussed below. The knowledge of Eq. (\ref{Cab}) is sufficient to derive the following scaling relations
\begin{equation}
\left\langle \left(s_{a}-\displaystyle\frac{1}{N}\sum_b s_{ab}\right)^2\right\rangle\propto \left\langle \left(s-\displaystyle\frac{1}{N}\sum_b s_{a}\right)^2\right\rangle\propto \frac{\alpha-1}{N}
\label{varsa}
\end{equation}
where $\alpha=(1/N)\sum_a \rho_a^2$ is a measure of non-equivalence of the channels. Using $\rho_a$ or ${\tilde \rho_a}$ gives $\alpha>1$, Table \ref{table}. As one can see, the breakdown of the composition rules Eq. (\ref{sa}) is in itself a mesoscopic effect related to the channel dispersion. Eq. (\ref{varsa}) also shows that the normalization in the definitions of $s_{ab}$ and $s_a$ is sufficient to remove the channel dependence in studies of mesoscopic fluctuations in $N\rightarrow\infty$ limit.

\begin{center}
\begin{table}
\begin{tabular}{lccc}
\hline
  & $\alpha_\rho-1$                     & $\alpha_{\tilde \rho}-1$                          &  $\alpha_{{\tilde \rho}_{d}}-1$ \\
\hline
2D& $0.022$ & $\displaystyle\frac{2/3+\pi^2/16}{(\Delta+\pi/4)^2}\approx 0.019$ & $\displaystyle\frac{2/3+\pi^2/16}{(\Delta_{d}+\pi/4)^2}\approx 0.020$ \\
3D& $0.035$ & $\displaystyle\frac{1/18        }{(\Delta+  2/3)^2}\approx 0.029$ & $\displaystyle\frac{1/18        }{(\Delta_{d}+  2/3)^2}\approx 0.031$ \\
\hline
  & $\beta_\rho$                        & $\beta_{\tilde \rho}$                             &  $\beta_{{\tilde \rho}_{d}}$ \\
\hline
2D& $\displaystyle\frac{2\sqrt{2}}{\pi}\approx 0.900$ & $\displaystyle\frac{\Delta+2/\pi}{\Delta+\pi/4}\approx 0.907$  & $\displaystyle\frac{\Delta_{d}+2/\pi}{\Delta_{d}+\pi/4}\approx 0.905$\\
3D& $\displaystyle\frac{\sqrt{3}}{2}   \approx 0.866$ & $\displaystyle\frac{\Delta+1/2}{\Delta+2/3}    \approx 0.879$  & $\displaystyle\frac{\Delta_{d}+1/2}{\Delta_{d}+2/3}    \approx 0.875$\\
\hline
\end{tabular}
\caption{\label{table} $\alpha$ and $\beta$ defined in Eqs. (\ref{varsa},\ref{beta}) quantify the deviation from ECA. The values in three columns were computed in $N\rightarrow\infty$ limit, where $\sum_a$ can be replaced with the appropriate integration. $\rho(\mu)$ from Eq. (\ref{rho}) was used in the first column. Approximation ${\tilde \rho}(\mu,\Delta)$ with $\Delta$ from RTT and diffusion theory were used to compute the values in the second and the third columns respectively. }
\end{table}
\end{center}

\begin{figure}
\vskip -0.5cm
\centerline{\rotatebox{-90}{\scalebox{0.25}{\includegraphics{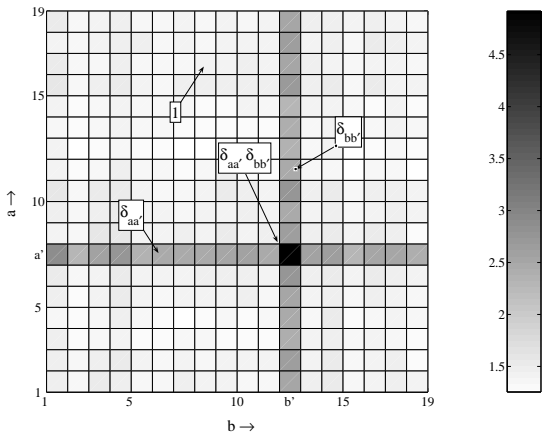}}}}
\vskip -0.5cm
\caption{\label{fig_Cab} Channel intensity correlations $\langle T_{ab}T_{a^\prime b^\prime}\rangle$ normalized by $\langle T_{ab}\rangle$ and $\langle T_{a^\prime b^\prime}\rangle$ to remove dependence on $\rho_{a,b}$. General structure of the resultant correlator remains the same in all studied samples, regardless of $g$. Sample with $g=0.4,\ N=19$ is shown in the figure. To visualize the four-dimensional array $a^\prime =7,\ b^\prime=12$ was chosen. }
\end{figure}

{\it Channel-channel correlations:} 
Diagrammatic perturbative calculations of the channel correlations in the slab geometry \cite{feng_kane_lee_stone,feng_phys_rep,rossum_rmp} give a general relation $\langle T_{ab}T_{a^\prime b^\prime}\rangle\propto \langle T_{ab}\rangle \langle T_{a^\prime b^\prime}\rangle$. Here $a,b$ denote different transverse momenta. The averages $\langle T_{ab}\rangle$ appear when the incoming and outgoing paths are paired into diffusons \cite{rossum_rmp}, the only contributions that survive averaging over disorder. The fact that $\langle T_{ab}\rangle$ has nontrivial dependence on $a,b$ in Q1D has not been widely appreciated. It is the lack of this dependence that makes (nonperturbative) RMT result \cite{mello_correlations} not directly applicable. Comparison with the results of the previous section shows that modifications needed to correct this shortcoming may not be straightforward. 

We find channel-resolved field and the intensity correlations in the form (Fig. \ref{fig_Cab})
\begin{eqnarray}
&\langle t_{ab}t_{a^\prime b^\prime}^*\rangle=\displaystyle\frac{\langle T\rangle\rho_a\rho_b}{N^2}\delta_{aa^\prime}\delta_{bb^\prime};
\ \ \  \langle T_{ab}T_{a^\prime b^\prime}\rangle =                                             \label{Ceab}\\
&\displaystyle\frac{\langle T\rangle^2\rho_a\rho_b\rho_{a^\prime}\rho_{b^\prime}}{N^4}
\times\left[ A\delta_{aa^\prime}\delta_{bb^\prime}+B(\delta_{aa^\prime}+\delta_{bb^\prime})+A\right] \label{Cab}
\end{eqnarray}
where $A,B$ are some functions of $g$. The structure of the above expression coincides with RMT prediction \cite{mello_correlations} only in ECA,  {\it i.e.} by setting $\rho_a\equiv 1$. Therefore, multiplication by $\rho_a\rho_b\rho_{a^\prime}\rho_{b^\prime}$ is sufficient to recover the proper correlations function, which does describe the diffusion within the crossection of the waveguide. It is also consistent with the expressions obtained with diagrammatic techniques in the slab geometry \cite{feng_kane_lee_stone,feng_phys_rep,rossum_rmp} where $\rho_{a,b}$ become continuous function of $\mu$. Upon multiplication by $\rho_{a,b}$'s, the correlations Eqs. (\ref{Ceab},\ref{Cab}) regain the dependence on the waveguide dimensionality and its shape (in 3D) that is in contrast to RMT where only number of channels enters as a sole parameter.

\begin{figure}
\vskip -0.3cm
\centerline{\rotatebox{0}{\scalebox{0.18}{\includegraphics{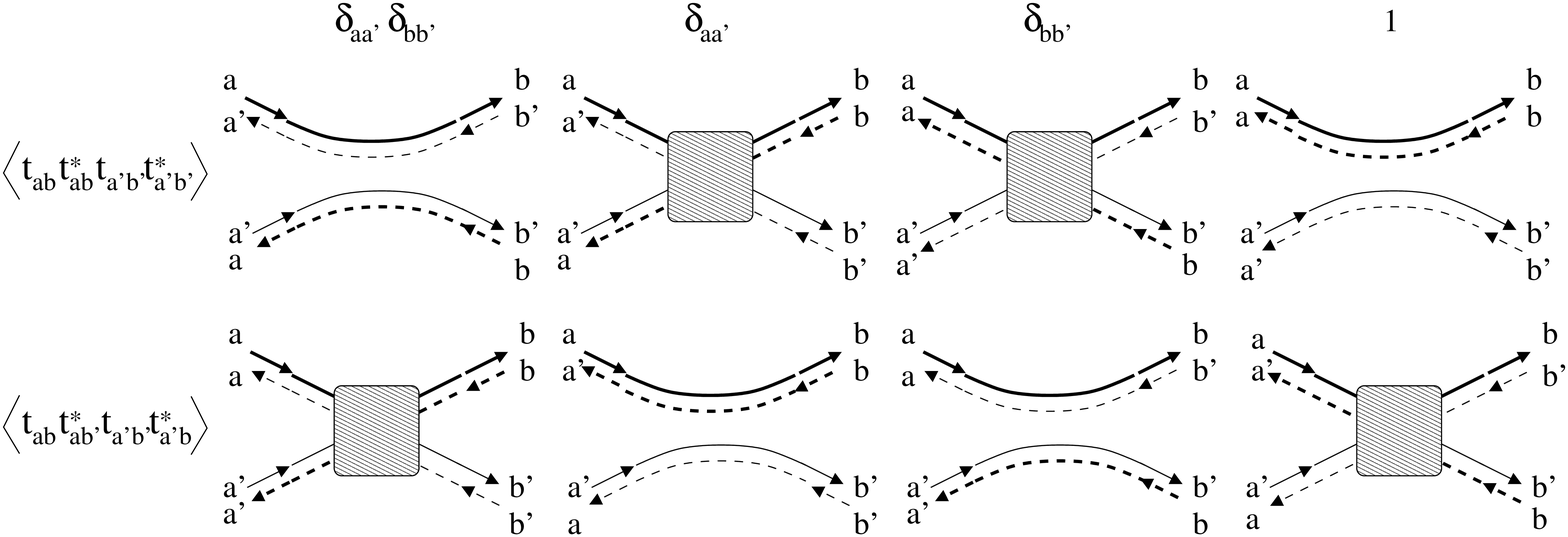}}}}
\vskip -0.2cm
\caption{\label{diagrams} Two types of channel correlators and their leading diagrams in $1/g$ perturbation series that appear in derivation of spacial intensity correlation function, Eq. (\ref{C}). The known expression for $\langle T_{ab}T_{a^\prime b^\prime}\rangle$ is shown in the first line, see {\it e.g.} \cite{rossum_rmp}. Leading contributions to the correlator $\langle t_{ab}t^{*}_{ab^\prime }t_{a^\prime b^\prime }t^{*}_{a^\prime b}\rangle$ can be obtained by swapping the incoming and outgoing indexes as shown in the second row, Eq. (\ref{Cabnew}) follows.}
\vskip -0.5cm
\end{figure}

{\it Spacial correlations:} 
Spacial correlations of the transmitted fields and intensities in Q1D have been a focus of the experimental and numerical studies \cite{azi_field_corr_fun,azi_building_block,corr_passive,yamilov_intensity_corr}. The results were compared to diagrammatic calculations in slabs. Despite the agreement with the theory, this may not be well justified because in Q1D (where the waveguide width $W$ is smaller then $L$) the boundary conditions in the transverse direction(s) differ from those in a slab. At the same time, starting with the RMT's (which enforces proper boundary conditions through quantization of the transverse momentum) expectation $\langle t_{ab}t^{*}_{a^\prime b^\prime}\rangle=(\langle T\rangle /N^2)\delta_{aa^\prime}\delta_{bb^\prime}$, accounting for the variation of the propagation speed for different modes and then averaging over the output cross-section of the waveguide gives the following expression (2D) $C_E(\Delta y)=\langle E(y_0+\Delta y)E^*(y_0)\rangle\propto\Sigma_{b=1}^N (1/k_b)\cos(\pi b\Delta y/W)$. It does not agree neither with our simulation nor diffusion theory\cite{freund_surf_corr_fun}, inset in Fig. \ref{fig_rho}, and leads to unphysical non-zero correlations $\propto 1/N$ for large separations.

We will see that account of the channel dependence in $\rho_a$'s corrects the discrepancy. In case of intensity correlations, the starting point in making transformation from channel to spacial coordinates is the correlator $\langle t_{a_1b_1}t^{*}_{a_2b_2}t_{a_3b_3}t^{*}_{a_4b_4}\rangle$. By pairing the channel indexes into diffuson contributions one can see that this correlator contains four terms 
$\delta_{a_1a_2}\delta_{b_1b_2}\delta_{a_3a_4}\delta_{b_3b_4}+
 \delta_{a_1a_4}\delta_{b_1b_4}\delta_{a_2a_3}\delta_{b_2b_3}+
 \delta_{a_1a_2}\delta_{b_1b_4}\delta_{a_3a_4}\delta_{b_2b_3}+
 \delta_{a_1a_4}\delta_{b_1b_2}\delta_{a_2a_3}\delta_{b_3b_4}$. The first two lead to Eq. (\ref{Cab}), whereas the last two require the knowledge of the new correlator $\langle t_{ab}t^{*}_{ab^\prime }t_{a^\prime b^\prime }t^{*}_{a^\prime b}\rangle$, which according to our diagrammatic analysis, Fig. \ref{diagrams},  has form:
\begin{equation}
\left(\langle T\rangle^2/N^4\right)\rho_a\rho_b\rho_{a^\prime}\rho_{b^\prime}
\times\left[ B\delta_{aa^\prime}\delta_{bb^\prime}+A(\delta_{aa^\prime}+\delta_{bb^\prime})+B\right] \label{Cabnew}
\end{equation}
with the same $A,B$ as in Eq. (\ref{Cab}). This is fully supported by our simulations for all studied samples. Based on Eqs. (\ref{Ceab},\ref{Cab},\ref{Cabnew}) we derive the following expressions for the spacial field and intensity correlation functions in a 2D waveguide:
\begin{widetext}
\begin{eqnarray}
\langle E(Y_0+\Delta Y,y_0+\Delta y)E^*(Y_0,y_0)\rangle&=& 
\displaystyle\frac{\langle T\rangle  \ \gamma^2}{k^2 W^2}\times f(\Delta Y)\; f(\Delta y); 
\ \ \ \ \ \ \ f(y)=\displaystyle\frac{1}{\gamma}\sum_a\frac{\rho_a}{\mu_a}\cos\left(\frac{\pi}{W}\;a\; y\right)                       \label{Ce}\\
\langle I(Y_0+\Delta Y,y_0+\Delta y)I  (Y_0,y_0)\rangle&=&
\displaystyle\frac{\langle T\rangle^2\ \gamma^4}{k^4 W^4}\times  \left\{A\; f^2(\Delta Y)\;f^2(\Delta y)+B\left[ f^2(\Delta Y)+f^2(\Delta y)\right]+A\right\} \label{C} 
\end{eqnarray}
\end{widetext}
where $\gamma[\rho_a]=\sum_a\rho_a/\mu_a$ and $Y,y$ are respectively the transverse position of the source and detector in the perfect lead regions before and after the disordered part of the waveguide. In the above, averaging over $Y_0,y_0$ where performed and only $O(1/N^0)$ terms were retained. The derived Eq. (\ref{Ce}) gives an excellent fit to data, inset in Fig. \ref{fig_rho}. Furthermore, substituting ${\tilde \rho}(\mu,\Delta_d)$ and changing summation over $\mu_a$ to the integral over $\mu$, it gives exactly the diffusion expression for 2D slabs \cite{freund_surf_corr_fun,corr_passive}. The structure of Eq. (\ref{C}) also agrees with the perturbative expression \cite{azi_building_block} where $A=1+2/15g^2$ and $B=2/3g$.

Setting $\Delta Y$ and $\Delta y$ to zero in Eqs. (\ref{Ce},\ref{C}) gives the first two moments of the transmitted intensity. Their analysis shows that channel dependence in $\rho_a$ also affects the actual value of the {\it average intensity}. Indeed, RMT's expression not corrected by $\rho_a$ would be reduced by a factor 
\begin{equation}
\beta^2=
\displaystyle\left( \frac{\gamma[\rho_a]}{\gamma[\rho_a\equiv 1]}                 \right)^2=
\displaystyle\left( \sum_a\frac{\rho_a}{\mu_a}/\displaystyle\sum_a\frac{1}{\mu_a} \right)^2.
\label{beta}
\end{equation}
which is computed Table \ref{table} as $\beta<1$. 

{\it Conclusion:}
The following two observations provide an insight into why simple multiplicative correction by $\rho_a$ was able to successfully remedy the limitation of RMT set by ECA: (i) The function $\rho_a$ is independent of system size, and thus is insensitive of the nature of wave transport inside random medium; (ii) spacial field correlation function predicted by RMT in 2D (the same conclusion holds also in 3D) gives Bessel function $C_E(\Delta y)\propto\Sigma_{b=1}^N (1/k_b)\cos(\pi b\Delta y/W)=J_0(ky)$ which coincides with Shapiro \cite{shapiro_ce} expression for the correlations {\it inside} random medium. Therefore, we conclude that $\rho_{a,b}$ describes the {\it surface effect}, whereas RMT properly describes the transport in the bulk of the random medium. Multiplication of RMT's transfer matrix by $\rho_{a,b}$ -- the escape function -- makes the {\it geometry dependent} boundary correction that leads to correct result in $N\rightarrow\infty$ limit. We derived Eqs. (\ref{Ceab},\ref{Cab},\ref{Cabnew},\ref{Ce},\ref{C}) based on the RMT with the surface correction that is consistent with known expressions for slab geometry. The obtained expressions also agree with our direct numerical simulations in both $g<1$ and $g>1$ regimes. Albeit the studied correlations are directly measurable only in optics, our results are also applicable to noninteracting electronic systems -- a volume-disordered wire, where time reversal symmetry is preserved and dephasing mechanisms can be neglected.

{\it Acknowledgements:}
The author is grateful to B. van Tiggelen, P. Sebbah and A. Z. Genack for valueable comments.

\end{document}